\documentclass[]{aa}
\usepackage{psfig}
\usepackage{times}
\usepackage{graphics}

\def\la{\;
\raise0.3ex\hbox{$<$\kern-0.75em\raise-1.1ex\hbox{$\sim$}}\; }
\def\ga{\;
\raise0.3ex\hbox{$>$\kern-0.75em\raise-1.1ex\hbox{$\sim$}}\; }
 
\newcommand{\zabs}{$z_{\rm abs}\,$}
\newcommand{\zem}{$z_{\rm em}\,$}

\newcommand{\kms}{km~s$^{-1}\,$}
\newcommand{\cm}{cm$^{-2}\,$}
\newcommand{\cmm}{cm$^{-3}\,$}

\begin{document}

\title{\object{HE 0141--3932}: a bright QSO with an unusual emission line spectrum 
and associated absorption\thanks{Based on observations obtained at
the VLT Kueyen telescope and 3.6~m ESO telescope, Chile, the ESO
programs No.~67.A-0280(A) and 72.D-0174A, respectively.}}
\author {D. Reimers\inst{1}
\and E. Janknecht\inst{1}
\and C. Fechner\inst{1}
\and I. I. Agafonova\inst{2}
\and S. A. Levshakov\inst{2}
\and S. Lopez\inst{3}}

\institute{Hamburger Sternwarte, Universit\"at Hamburg, 
Gojenbergsweg 112, D-21029 Hamburg, Germany
\and Department of Theoretical Astrophysics, 
Ioffe Physico-Technical Institute, 194021 St. Petersburg, Russia
\and Departamento de Astronomia, Universidad de Chile, 
Casilla 36-D, Santiago, Chile}
\offprints{S. A.~Levshakov, \protect \\lev@astro.ioffe.rssi.ru}
\date{received date; accepted date}

\abstract{\object{HE 0141--3932} (\zem = 1.80)
is a bright blue radio-quite quasar with
an unusually weak Ly$\alpha$ emission line. 
Large redshift differences ($\Delta z = 0.05$) are observed
between high ionization and low ionization emission lines.
Absorption systems identified at \zabs = 1.78, 1.71, and 1.68
show mild oversolar metallicities ($Z \approx 1-2Z_\odot$) and can be attributed
to the associated gas clouds ejected from the circumnuclear region.
The joint analysis of the emission and absorption lines leads to the
conclusion that this quasar is seen almost pole-on.
Its apparent luminosity may be Doppler boosted by $\sim 10$ times.
The absorbing gas shows a high abundance of Fe, Mg and Al\, 
([Fe, Mg, Al/C] $\simeq 0.15\pm0.10$)
along with underabundance of N\, ([N/C] $\leq -0.5$). 
This abundance pattern is at variance with current chemical evolution models
of QSOs predicting [N/C] $\ga 0$ and [Fe/C] $<0$
at $Z \sim Z_\odot$.
} 
\authorrunning{D. Reimers et al.}
\titlerunning{\object{HE 0141--3932}: a bright QSO with an unusual
emission line spectrum and associated absorption}
\maketitle

\begin{figure*}[t]
\vspace{0.0cm}
\hspace{0.0cm}\psfig{figure=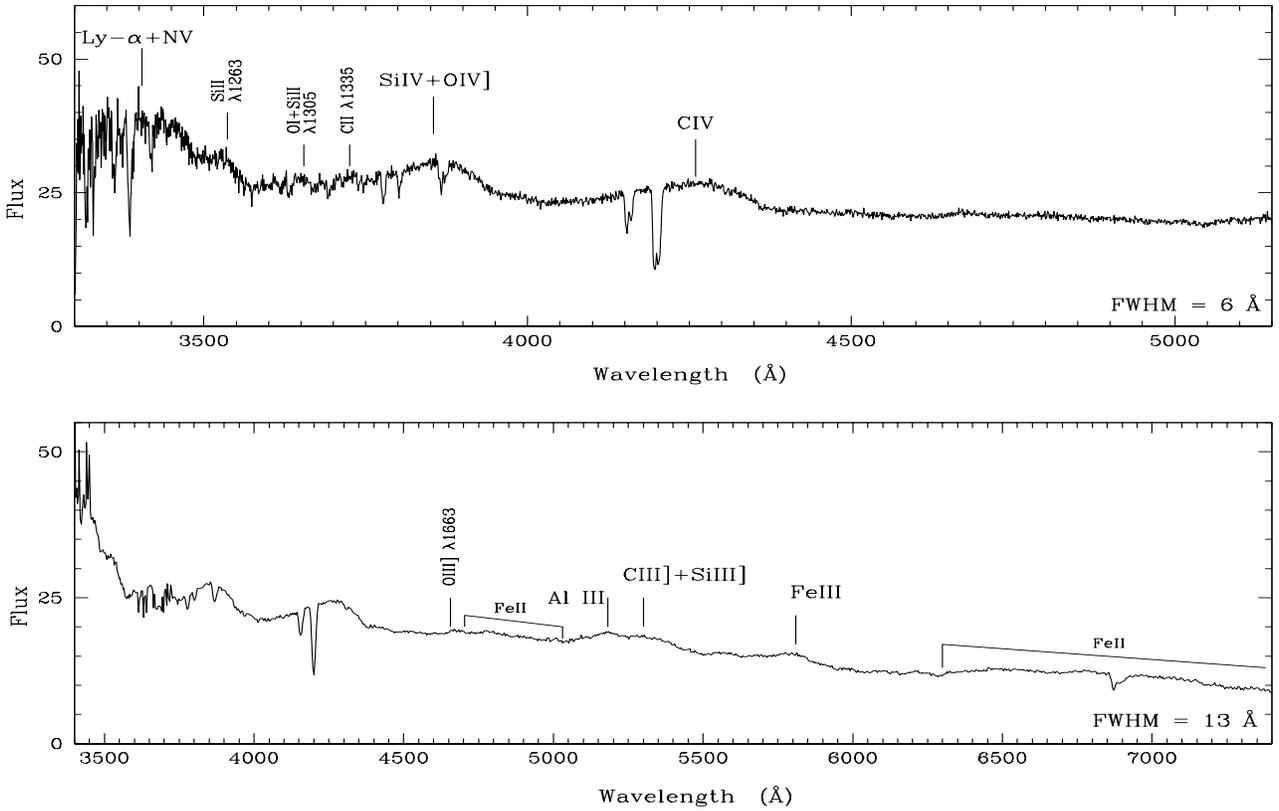,height=14.0cm,width=18.0cm}
\vspace{-3.0cm}
\caption[]{EFOSC~2 spectra of \object{HE 0141--3932}. 
The spectral resolutions are indicated in the panels. 
All identified emission features are labeled. The parameters of the
prominent emission lines are listed in Table~2
}
\label{fig1}
\end{figure*}

\section{Introduction}

In the course of a high-resolution study of the Ly$\alpha$ forest at 
intermediate redshifts (1.5 $\leq z \leq$ 2) in bright quasars
from the Hamburg/ESO Survey with 
the UV-visual echelle spectrograph (UVES) at the VLT, the 
discovery spectrum of the QSO 
\object{HE 0141--3932} ($z \simeq$ 1.8, $B = 16.2$, Wisotzki et al. 2000)
attracted our attention by two facts: it appeared to have no or only very 
weak Ly$\alpha$ emission along with clearly recognizable 
\ion{Mg}{ii}, and it showed several 
absorption line systems at \zabs ranging from 1.78 to 1.68 with very complex 
and strong metal profiles which allow us to
suggest that these systems may originate in the ejected gas. 

Since the identification spectrum of \object{HE 0141--3932} was of only moderate 
quality, we took further low-resolution spectra with EFOSC~2 with the
ESO 3.6~m telescope to improve the redshift measurement and 
we found a third peculiarity, namely large redshift differences
($\Delta z \sim 0.05$) between high ionization and low ionization emission 
lines. 

Emission line spectra similar to that of \object{HE 0141--3932}
are observed  in some high redshift BL Lacertae (BL Lac) objects 
(Urry \& Padovani 1995, hereafter UP95).
However, \object{HE 0141--3932} is a radio-quiet quasar, i.e. not a blazar.
A few other blue radio-quiet quasars are known with
apparently missing or weak Ly$\alpha$ emission but clearly present
metal lines: 
\object{PG 1407+265} [\zem = 0.94, McDowell et al. (1995)], 
\object{Tol 1037--2703} [\zem = 2.20, Srianand \& Petitjean, (2001)], 
\object{PHL 1811} [\zem = 0.192, Leighly, Halpern, \& Jenkins (2004)],
and several  cases from SDSS (Fan et al. 1999, 2003). 
The nature of their peculiar emission line spectra is not clear.

In particular, Leighly et al. (2004) suggest a high accretion rate which
powers the UV emission from an optically thick
accretion disk, while suppressing the formation of a hot corona.
However, recent radio observations of \object{PG 1407+265}
on the milliarcsecond scale revealed
a relativistic jet of moderate power beamed toward us 
(Blundell, Beasley \& Bicknell 2003).
This means that the quasar is seen almost pole-on and its emission line spectrum
may be diluted by continuum radiation both from the jet and the accretion disk.

In the present paper we analyze both the emission and absorption 
line spectra of \object{HE 0141--3932}.   
We expect that this joint consideration may shed some light on the unusual
properties of this quasar.
  
The paper is organized as follows:
observations are described in Sect.~2,
the emission lines are studied in Sect.~3, the analysis of  
the \zabs $\approx$ \zem absorbers is given in Sect.~4, 
the results are discussed and summarized in Sect.~5.

\section{Observations}

\object{HE 0141--3932} was observed with the UVES 
at the 8~m ESO VLT on Paranal over 7 nights in July/August 2001.
Eleven individual exposures with integration times of 60 min were made 
using the dichroic mode in standard settings (Table~1). With a slit
width of $1''$, a resolution of 41000 (7.3 \kms) in the blue and 38000
 (7.9 \kms) in the red is achieved.

The data reduction was performed at the Quality Control Group in Garching
using the UVES pipeline Data Reduction Software (Ballester et al. 2000), 
the vacuum-barycentric corrected spectra were co-added.
The resulting signal-to-noise ratio, S/N, is typically 75.

Since both the Ly$\alpha$ and the \ion{Mg}{ii} emission line ranges 
are covered only by UVES spectra, we performed an absolute flux calibration 
using an appropriate master response curve and the respective airmasses 
during the observations. The procedure is described on the ESO web page
www.eso.org/observing/dfo/quality/UVES/qc/response.html. 

Further spectra were taken with EFOSC~2 at the ESO 3.6~m telescope 
on October 2, 2003, to improve our knowledge
about the redshifts of the emission lines. Details are given in Table 1. 
Wavelength and flux calibration was performed according to standard
procedures. The resulting spectra are shown in Fig.~1. 
Notice that due to the rapidly decreasing sensitivity below 3500 \AA, 
the EFOSC~2 spectra only partially cover the wavelength range expected for 
the Ly$\alpha$ emission line.

\section{The emission line spectrum}

The combination of EFOSC~2 spectra with roughly flux calibrated 
UVES spectra allows us to estimate redshifts
and equivalent widths of all emission lines between
Ly$\alpha$ and \ion{Mg}{ii}~$\lambda2800$ \AA. 
Equivalent widths ($EW$) and redshifts were measured by fitting 
Gaussian profiles to the data.
Table~2 presents the results. 
The Ly$\alpha$ $EW$ is relatively uncertain since in the calibrated
UVES data the continuum is difficult to determine and the blue 
wing of the line is incomplete (see Fig.~1 and 2).

The identification of the emission line at $\sim$ 3400 \AA\, is ambiguous: 
it could be either Ly$\alpha$ at $z = 1.80$ or \ion{N}{v} at
$z = 1.75$ or blend of both lines. 
There are two arguments in favor of the \ion{N}{v} identification: 
it has a width comparable to \ion{C}{iv} and its redshift -- if identified
as \ion{N}{v} --  is equal to that of the \ion{C}{iv},  
while in QSOs with significant redshift differences between high
ionization lines (\ion{C}{iv}) and low ionization lines 
(\ion{Mg}{ii}, H$_{\alpha}$), the Ly$\alpha$ line typically is found 
at the redshift of the
high ionization BLR emission lines (Gaskell, 1982; Espey et al., 1989). 
On the other hand, the emission line at 3400 \AA\, is quite strong for 
\ion{N}{v}, although there are quasars known with 
\ion{N}{v} strength comparable to that of \ion{C}{iv} (Hall et al. 2004;
Baldwin et al. 2003b). 
Even if this line would be entirely due to emission of neutral hydrogen, 
we can conclude that Ly$\alpha$ in \object{HE 0141--3932} is unusually weak.

\begin{table}[t]
\centering
\caption{Log of spectroscopic observations of \object{HE 0141--3932}}
\label{tab-1}
\begin{tabular}{lr@{--}lcll}
\hline
\noalign{\smallskip}
Instrument & \multicolumn{2}{c}{$\lambda\lambda$} & Exposure & Date & Resolution\\
           &  \multicolumn{2}{c}{ (\AA) }         & time (h) &      &  (\kms)\\
\noalign{\smallskip}
\hline
\noalign{\smallskip}
UVES  & 3053&3872  & 7   &         & 7.3 \\
      & 4700&5800  & 7   & July +  & 7.3 \\
      & 5800&6800  & 7   & August  & 7.8 \\
UVES  & 3761&4982  & 4   & 2001    & 7.3 \\
      & 6650&8550  & 4   &         & 7.8 \\
      & 8600&10400 & 4   &         & 7.8 \\
EFOSC~2 & 3270&5240  & 0.25 & Oct. 2 & 420\, (6\,\AA)\\
      & 3380&7520  & 0.25 & 2003   & 700\, (13\,\AA)\\
\noalign{\smallskip}
\hline
\end{tabular}
\end{table}

The second anomaly of this quasar is the large velocity difference 
between low ionization lines 
(\ion{Mg}{ii}, \ion{C}{iii}], \ion{Al}{iii} etc.) and high
ionization lines 
(\ion{C}{iv}, \ion{Si}{iv}) of roughly 5000 \kms; 
it is among the most extreme cases like 
\object{PG 1407+265} (McDowell et al. 1995, 
$\Delta v_{\rm Mg\,II-C\,IV} \simeq 5000$ \kms) and
\object{Q 1309--056} (Espey et al. 1989, 
$\Delta v_{\rm Mg\,II-C\,IV} \simeq 4000$ \kms).

In Fig.~2 we display the emission line profiles on a velocity scale relative 
to $z = 1.80$. The redshift of the low ionization lines of
$z = 1.80$ appears to be roughly the systemic redshift, 
since the UVES spectrum shows that the Ly$\alpha$ forest starts 
at $z = 1.8086$.

While the velocity shift in the emission lines is extremely large, 
how unusual are the emission line intensities~?
This can be best seen by a comparison with the histogram of equivalent 
widths of the LBQS-quasars, a well selected sample
of QSO given by Francis (1993). 
The Ly$\alpha$, \ion{C}{iv}, \ion{C}{iii}] and 
\ion{Mg}{ii} equivalent widths are among the bottom 2\%, 1\%, 13\%
and 1.5\%, respectively, of the Francis (1993) distribution. 
At the same time,
the \ion{Al}{iii}~1856/64 and \ion{Fe}{iii}~2075 lines appear 
unusually strong.
In order to investigate the physical properties of gas
which could produce such an unusual emission line spectrum 
we compared the observed equivalent widths of \object{HE 0141--3932} 
with the compilation
of quasar BLR rest-wavelength emission line equivalent widths given 
as functions of column density, incident ionizing
spectrum, and metal abundance of the emitting gas clouds 
(Korista et al. 1997).
While Ly$\alpha$, \ion{Mg}{ii} and \ion{C}{iii}] 
have roughly the same velocity (redshift) and could originate in the
same volume, we were unable to find a parameter combination which met 
the constraints on Ly$\alpha$
and \ion{Mg}{ii} line strengths simultaneously -- 
the theoretical Ly$\alpha$ is always much too strong for a
parameter combination that matches the \ion{Mg}{ii} equivalent width.
The \ion{C}{iv}/Ly$\alpha$ $EW$ ratio can be met 
assuming high gas densities
[log~n(H) $\geq$ 12], but
the discrepant redshifts exclude formation in the same location.
A mean BLR parameter set appears to be unable to reproduce the 
measured $EW$s of emission lines in \object{HE 0141--3932}.

\begin{figure}[t]
\centering
\vspace{0.0cm}
\hspace{0.0cm}\psfig{figure=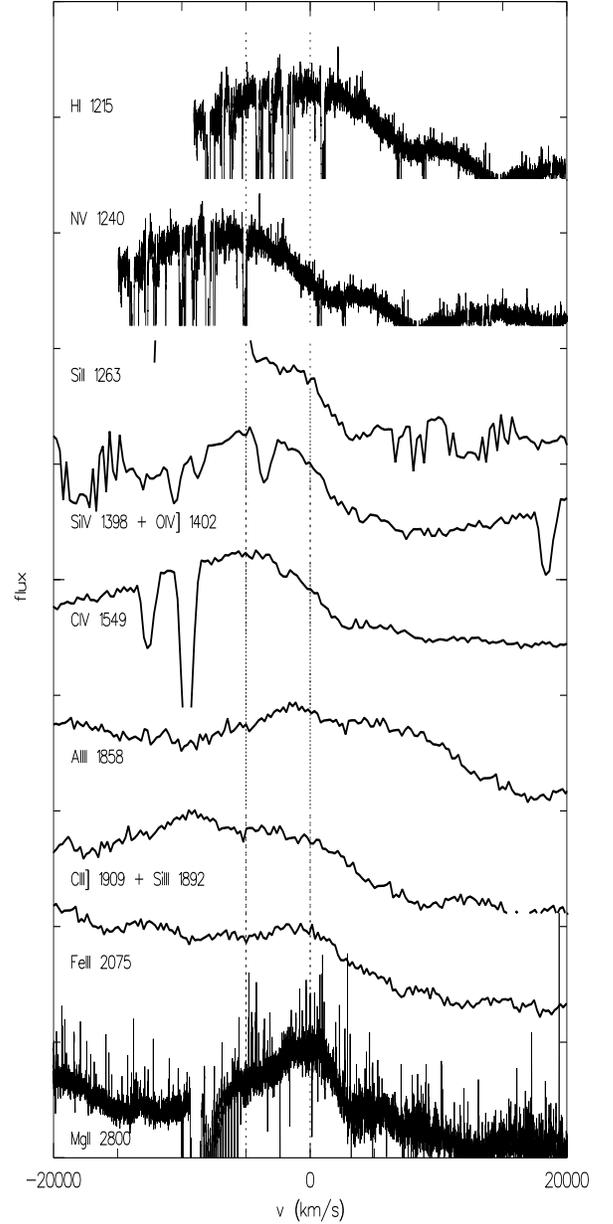,height=17.0cm,width=6.0cm}
\vspace{0.3cm}
\caption[]{Emission lines of \object{HE 0141--3932} (spectra from UVES and EFOSC~2).
$v = 0$ \kms\, corresponds to $z = 1.80$. For orientation, at $v = 0$ \kms\,
and $v = -5000$ \kms\, dashed lines are plotted
}
\label{fig2}
\end{figure}

\begin{table*}[t]
\caption{Estimated equivalent widths, redshifs and relative
velocities for emission lines of \object{HE 0141--3932}}
\label{tbl-t2}
\begin{tabular}{lcr@{ $\pm$ }lcr@{ $\pm$ }ll}
\hline
\noalign{\smallskip}
Ion & Instrument & 
\multicolumn{2}{c}{$EW_{\rm rest}$, \AA} 
& $z_{\rm obs}$ & 
\multicolumn{2}{c}{$v$, \kms} & Comment\\
\noalign{\smallskip}
\hline
\noalign{\smallskip}
\ion{H}{i} 1215&UVES&16&7&$1.80^{+0.01}_{-0.02}$&
\multicolumn{2}{c}{$0^{+1070}_{-2140}$} & asymmetric \\
and/or \ion{N}{v} 1239 & &
\multicolumn{2}{c}{ } & $1.75\pm0.01$ & $-5350$&1070& \\
\noalign{\smallskip}
\ion{Si}{iv} 1398+\ion{O}{iv}] 1402 & EFOSC~2&11&1&$1.76\pm0.01$&
$-4280$&1070 & --- \\
\noalign{\smallskip}
\ion{C}{iv} 1549 & EFOSC~2 &14&1 & $1.75\pm0.01$ &$-5350$&1070 
& asymmetric \\
\noalign{\smallskip}
\ion{Al}{iii} 1858 + & EFOSC~2 &13&1& $1.79\pm0.01$ &$-1070$&1070 & --- \\
\ion{C}{iii}] 1909 + \ion{Si}{iii}] 1892 & &
\multicolumn{2}{c}{ } &$1.79\pm0.01$ &$-1070$&1070 
& blend, fractions of components unknown\\
\noalign{\smallskip}
\ion{Fe}{iii} 2075 & EFOSC~2 & 5&1 & $1.80\pm0.01$ & $0$&1070 & --- \\
\noalign{\smallskip}
\ion{Mg}{ii} 2800 & UVES  & 16&1 & $1.795\pm0.005$ & $-535$&535
& blended with \ion{Fe}{ii} emission \\
\noalign{\smallskip}
\hline
\noalign{\smallskip}
\multicolumn{8}{l}{Zero velocity corresponds to $z = 1.80$. Only 
statistical errors are indicated.} 
\end{tabular}
\end{table*}

\section{The absorption systems}

\subsection{Calculation procedure}

In order to estimate the physical parameters of the absorption systems 
we used the Monte Carlo Inversion (MCI) method. 
Detailed description of the MCI is given in 
Levshakov et al. (2000, 2002, 2003a). 
Here we outline briefly its basics
needed to understand the results presented.
 
The MCI is based on the assumption
that all lines observed in the absorption system are formed in
a continuous absorbing
gas slab of thickness $L$ 
where the gas density, $n_{\rm H}(x)$, and velocity, $v(x)$,
fluctuate from point to point giving rise to complex profiles
(here $x$ is the space coordinate along the line of sight).
 
We also assume
that within the absorber the metal abundances are constant,
the gas is optically thin for the ionizing UV radiation, and the gas
is in thermal and ionization equilibrium.
The intensity and the spectral shape of the background ionizing
radiation are treated as external parameters.
 
The radial velocity  $v(x)$ and gas density $n_{\rm H}(x)$
are considered as two continuous random functions which are
represented by their sampled values at equally spaced intervals
$\Delta x$. The computational procedure is based on 
adaptive simulated annealing. The fractional ionizations of
different elements at each space coordinate $x$, 
$\Upsilon_{\rm ion}[U(x)]$,\,
are computed 
with the photoionization code CLOUDY (Ferland 1997).
 
The following physical
parameters are directly estimated by the MCI procedure:
the mean ionization parameter $U_0$,
the total hydrogen column density $N_{\rm H}$,
the line-of-sight velocity dispersion, $\sigma_{\rm v}$, and
density dispersion, $\sigma_{\rm y}$, of the bulk material
[$y \equiv n_{\rm H}(x)/n_0$],
and the chemical abundances $Z_{\rm a}$ of all elements
involved in the analysis.
With these parameters we can further calculate
the column densities for different species $N_{\rm a}$,
and the mean kinetic temperature
$T_{\rm kin}$.
 
In general, the uncertainties on the fitting parameters $U_0$, $N_{\rm H}$,
$\sigma_v$, $\sigma_y$, and $Z_a$ are about 15\%--20\%
(for data with S/N $\ga 30$)
and the errors of the estimated column densities are less than 10\%. 
However, in individual absorption systems 
the accuracy of the recovered values can
be lower for different reasons like partial blending of the line profiles,
saturation, or absence of lines of subsequent ionic transitions.

The MCI can be supplemented with an additional procedure 
aimed at restoring the shape of
the background ionizing spectrum. A formal
description of this procedure is given in the Appendix. 
Its accuracy depends significantly on the number of  
unsaturated lines of the subsequent
ionic transitions of different elements 
(e.g. \ion{Si}{ii}/\ion{Si}{iii}/\ion{Si}{iv}, 
\ion{C}{ii}/\ion{C}{iii}/\ion{C}{iv}) available for the analysis. 
The absorber at \zabs = 1.7817 (described below) 
reveals enough such lines and we use them to estimate
the spectral shape of the background UV radiation in the range $E > 1$ Ryd.

All calculations are carried out with the laboratory wavelengths 
and oscillator
strengths taken from Morton (2003). Solar photospheric abundance for 
carbon is taken
from Allende Prieto et al. (2002), for silicon, nitrogen and iron --
from Holweger (2001).

\begin{figure*}[t]
\vspace{0.0cm}
\hspace{-0.5cm}\psfig{figure=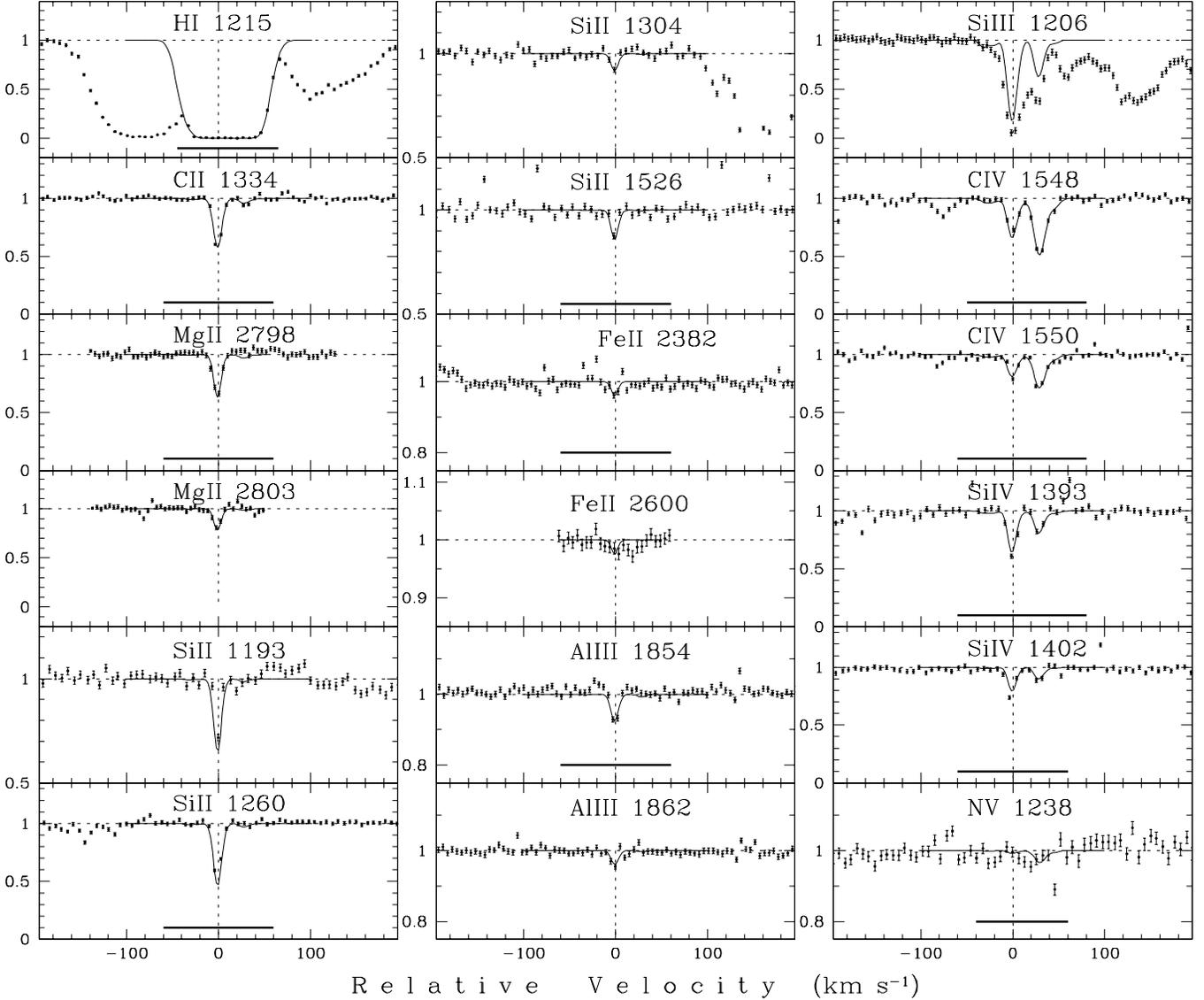,height=18.0cm,width=19.0cm}
\vspace{-2.8cm}
\caption[]{
Hydrogen and metal absorption lines associated with the \zabs = 1.7817 system
toward \object{HE 0141--3932} 
(normalized intensities are shown by dots with 1 $\sigma$
error bars). The zero radial velocity is fixed at $z = 1.7817$. Smooth curves
are the synthetic spectra convolved with the corresponding point-spread
spectrograph function and computed with the physical parameters listed in 
Table~3, Col.~2.
Bold horizontal lines mark pixels included in the optimization procedure.
The synthetic profiles of unmarked absorption features
were calculated in a second round using the
velocity $v(x)$ and gas density $n_{\rm H}(x)$
distributions already obtained in the optimization procedure.
}
\label{fig3}
\end{figure*}

\subsection{Absorption system at \zabs = 1.7817}

Many unsaturated lines of different ionic transitions along with a
single saturated hydrogen line Ly$\alpha$ are detected in this system (Fig.~3)
separated by 2\,000 km/s from the central source.
The blue wing of Ly$\alpha$ is partially blended by the adjacent 
hydrogen line [$N$(\ion{H}{i}) $\sim 10^{14}$ \cm] from an
intervening absorber (only a weak \ion{C}{iv}
doublet is detected in this system). 
\ion{C}{iii}~$\lambda977$ is 
beyond the wavelength coverage and  \ion{Si}{iii}~$\lambda1206$ 
is blended, probably
with  Ly$\alpha$  absorption from the \zabs = 1.7606. 

We started the analysis assuming  standard ionizing backgrounds such as a
power law $f_\nu \propto \nu^{-\alpha}$ 
(with different indexes $\alpha = 1.0 - 1.8$), 
the mean intergalactic spectrum (at $z = 1.8$) 
of Haardt \& Madau (1996), 
and the AGN-type spectrum of Mathews \& Ferland (1987, hereafter MF). 

None of these spectra was consistent with the observed intensities of
the \ion{C}{ii}, \ion{C}{iv}, \ion{Si}{ii} and \ion{Si}{iv} lines: all 
trials underproduced \ion{C}{ii}/\ion{C}{iv} and overproduced 
\ion{Si}{ii}/\ion{Si}{iv},
but the MF spectrum provided the lowest $\chi^2$. 
All spectra gave the mean ionization parameter in the range of 
$-2.75 \la \log U_0 \la -2.5$. 
In spite of the saturation, the neutral hydrogen column density can be
estimated with a sufficiently high accuracy ($\sim$ 20\%) since the velocity
dispersion of gas is determined by numerous metal lines detected in this
system (the procedure to restore partly blended profiles is
described in Sect.~3 in Levshakov et al. 2003a).
All runs with different model spectra
showed  a rather high metallicity -- slighty above solar value, and
an almost solar ratio of Si/C.

\begin{figure}[t]
\vspace{0.0cm}
\hspace{0.0cm}\psfig{figure=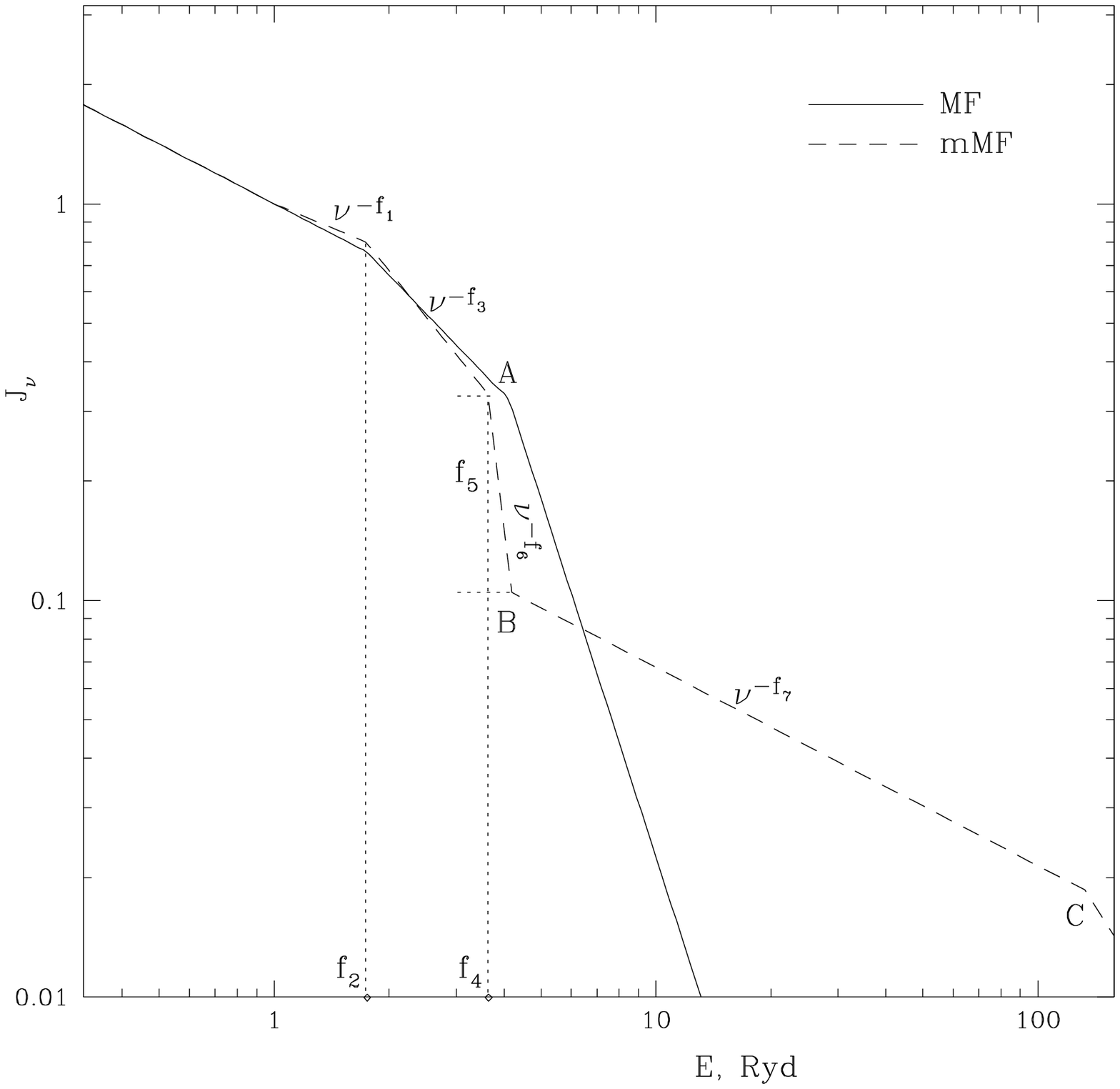,height=8.0cm,width=8.0cm}
\vspace{-0.7cm}
\caption[]{Spectral
shape of a typical AGN 
ionizing continuum from Mathews \& Ferland (1987) shown by the solid line
and its modification (dashed line) estimated for the \zabs = 1.7817 system. 
The spectra are normalized so that $J_\nu(h\nu =$ 1 Ryd) = 1.
Factors $f_i$ are defined in the Appendix
}
\label{fig4}
\end{figure}

Taking into account this preliminary information we can assume that the UV
spectrum to be found should maximize the product of ratios 
\ion{C}{ii}/\ion{C}{iv} and \ion{Si}{iv}/\ion{Si}{ii} 
in the range of $U_0 \sim 0.002 - 0.003$ for solar 
metal content and solar element abundances. 
The spectral shape of the MF spectrum can be taken as
a first approximation (for more details see Appendix). 

Applying the adjustment procedure as described in Appendix,
we estimated a new shape of the ionizing spectrum which ensured 
a self-consistent description of all lines observed in the 
\zabs = 1.7817 system. This spectrum
is shown in Fig.~4 by the dashed line, whereas the initial MF
spectrum is the solid line. As seen in Fig.~4, the
restored spectrum is softer in the range 2~Ryd $ < E < 6.5$ Ryd 
but much harder above 6.5 Ryd. 
It shows a break at the \ion{He}{ii} ionization edge with the
amplitude $J_\nu$(A)/$J_\nu$(B) $\simeq 3$
corresponding to the optical depth $\tau_c$(\ion{He}{ii}) $\simeq 1$.
Physical parameters estimated with this modified ionizing background are
listed in Table~3, Col.~2, and the corresponding synthetic profiles 
are plotted in Fig.~3 by the smooth curves.

\subsection{Absorption system at \zabs = 1.7103}

This absorption system spans the velocity range of 500 \kms\, and 
is located in velocity space at a distance of 9\,600 \kms\, from the QSO.
It reveals many lines of different ionic transitions (Fig.~5). 
The structure of the Ly$\alpha$, \ion{C}{iv}, \ion{Si}{iv} and \ion{N}{v} 
profiles indicates that the system may be divided into
two subsystems: one at
--200 \kms\,$< v <$ 180 \kms\, ($A$) and another at 
180 \kms\,$< v <$ 320 \kms\, ($B$).

The subsystem $B$ has an unsaturated hydrogen line and hence
allows us to estimate the hydrogen  column density 
with a sufficiently high accuracy.
The physical parameters computed  with the UV background deduced for the
\zabs = 1.7817 system are given in Table~3, Col.~4. 
The obtained extremely high metallicity 
-- almost $9 Z_\odot$ -- is striking. 
Since silicon lines \ion{Si}{iii} $\lambda1206$ and 
\ion{Si}{iv} $\lambda1393, 1402$ 
are very weak and the element ratios ([Si/C], [N/C]) not known a priori,
the mean ionization
parameter $U_0 \simeq 0.04$ represents a lower limit 
consistent with the observed intensities of the
\ion{C}{iv} lines and the absence of the \ion{C}{ii} $\lambda1334$ absorption.
Formally this subsystem could be fitted
with any higher ionization parameter producing even higher
metallicity and higher [Si/C] and [C/N] ratios.

To confirm the high metallicity in this subsystem 
we repeated calculations assuming other ionizing backgrounds -- power
laws with $\alpha$ ranging between  1.0 and 1.8 and the MF spectrum. 
These trials also delivered metallicity of
about 10 solar.

In principle,
such high metal enrichment of the circumnuclear gas is predicted 
in some models of chemical evolution of QSO/host galaxies
(Hamann \& Ferland 1999, hereafter HM).
However, another explanation of our results is that the absorbing
gas is not in equilibrium with the ionizing background, e.g.,
is still cooling and recombining.
In this case high metallicities can be caused by a longer recombination
time of hydrogen as compared to \ion{C}{iv} and \ion{N}{v}
(e.g., Osterbrock 1989).
The ionization parameter $U$ and, hence, the total hydrogen column density
$N_{\rm H} \propto 1/\Upsilon_{{\rm H\,}{\scriptscriptstyle\rm I}}(U)$
is determined by the observed 
line intensities of different ions.
Since hydrogen is ionized higher than it would be in the ionization equilibrium
with the same parameter $U$, the total hydrogen becomes underestimated
leading to an artificially high metallicity.
We may expect that the adjacent subsystem $A$ (centered at $v=0$ \kms) 
would help to clarify the
physical conditions in the \zabs = 1.7103 absorber.

The subsystem $A$ shows complex profiles of  low ionization
lines \ion{C}{ii}, \ion{Mg}{ii}, \ion{Si}{ii}, and
\ion{Fe}{ii} along with \ion{Al}{iii},  \ion{Si}{iii},
\ion{Si}{iv}, \ion{N}{v} and saturated \ion{C}{iv} (see Fig.~5).
The computations had been carried out with the ionizing background from the
\zabs=1.7817 system. The presence of the subsequent ionic transitions guarantees
the accuracy of the mean ionization parameter of $\sim 10$\% 
(for a given ionizing background).

The assumption of constant metallicity inside the
absorber turned out to be inconsistent with the observed blue
wing of Ly$\alpha$ and the absence of any metal absorption in the
range $-200$ \kms $< v <$ $-100$ \kms. 
A single available hydrogen line does not allow us 
to conclude whether the blue wing is blended or
there is indeed a gradient of metal content.
The former possibility looks more probable since
there is a non-zero flux in
the Ly$\alpha$ profile at --110 \kms $< v <$ --90 \kms\, 
with the mean intensity $0.018\pm0.005$ (marked by the arrow in Fig.~5).
The constant metallicity throughout this sub-system
seems to be appropriate since very steep blue wings of several
ions (\ion{Si}{iii}, \ion{Si}{iv},\ion{C}{iv}) do not indicate
a progressive dilution.
Thus, the calculations were carried out with the assumption of
constant metallicity. The neutral hydrogen column density and the shape of
the blue wing of the synthetic Ly$\alpha$ shown in Fig.~5
were calculated with the velocity
and density fields estimated from the observed metal line profiles
(for details see Sect.~4.3 in Levshakov et al. 2003a).
The obtained physical parameters are given in Table~3, Col.3.
We also tried 
other ionizing backgrounds (different power laws and the MF spectrum), but
none of them was consistent with the observed relative 
intensities within the \ion{C}{ii}/\ion{C}{iv}
and \ion{Si}{ii}/\ion{Si}{iii}/\ion{Si}{iv} profiles.

\begin{figure*}[t]
\vspace{0.0cm}
\hspace{-0.5cm}\psfig{figure=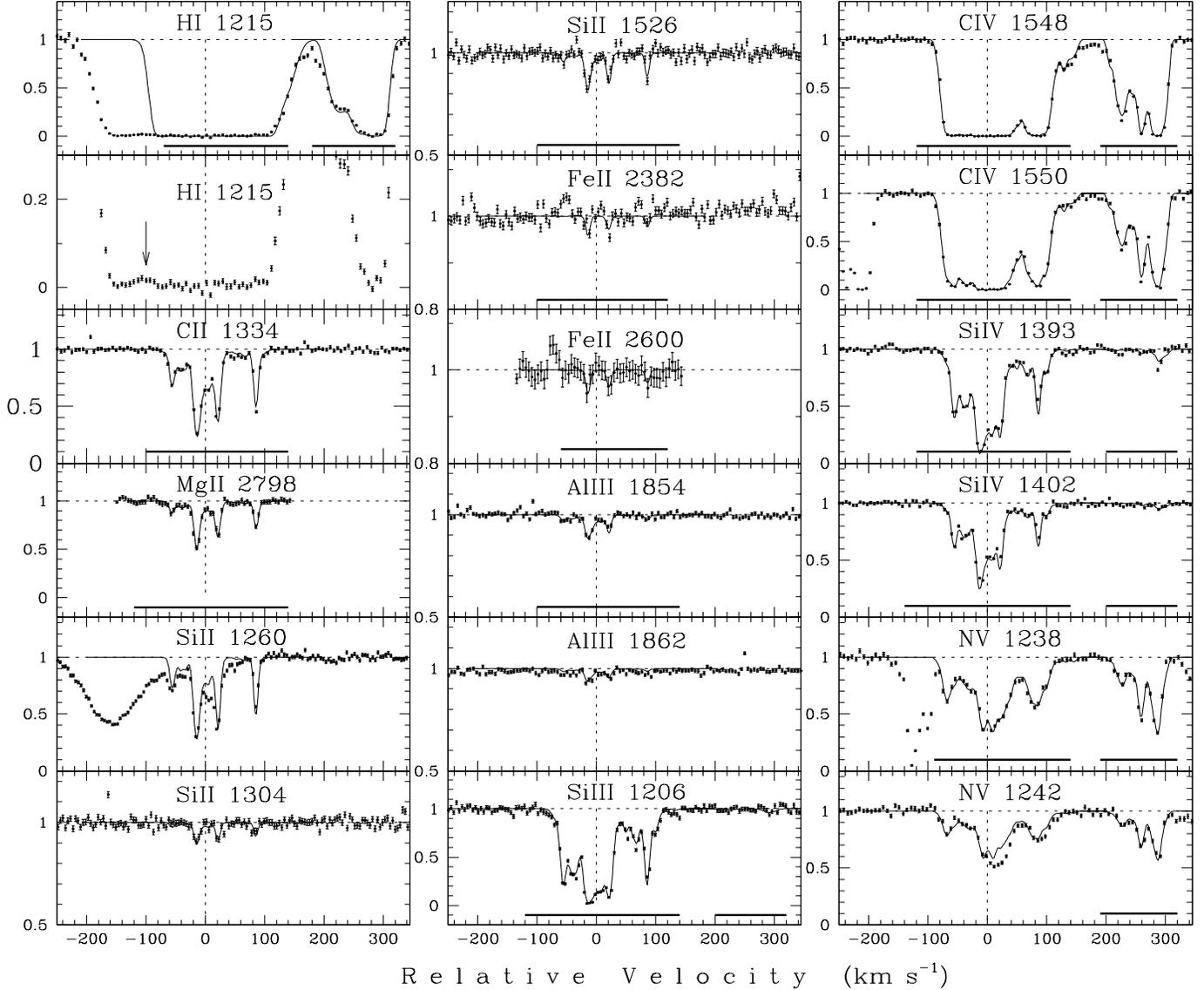,height=18.0cm,width=19.0cm}
\vspace{-3.0cm}
\caption[]{
Same as Fig.~1 but for the \zabs = 1.7103 system.
The zero radial velocity is fixed at $z = 1.71027$. 
The corresponding physical parameters are listed in Table~3, Cols.~3,4.
The arrow in the Ly$\alpha$ panel indicates pixels with non-zero intensities
}
\label{fig5}
\end{figure*}

\begin{figure*}[t]
\vspace{0.0cm}
\hspace{-0.5cm}\psfig{figure=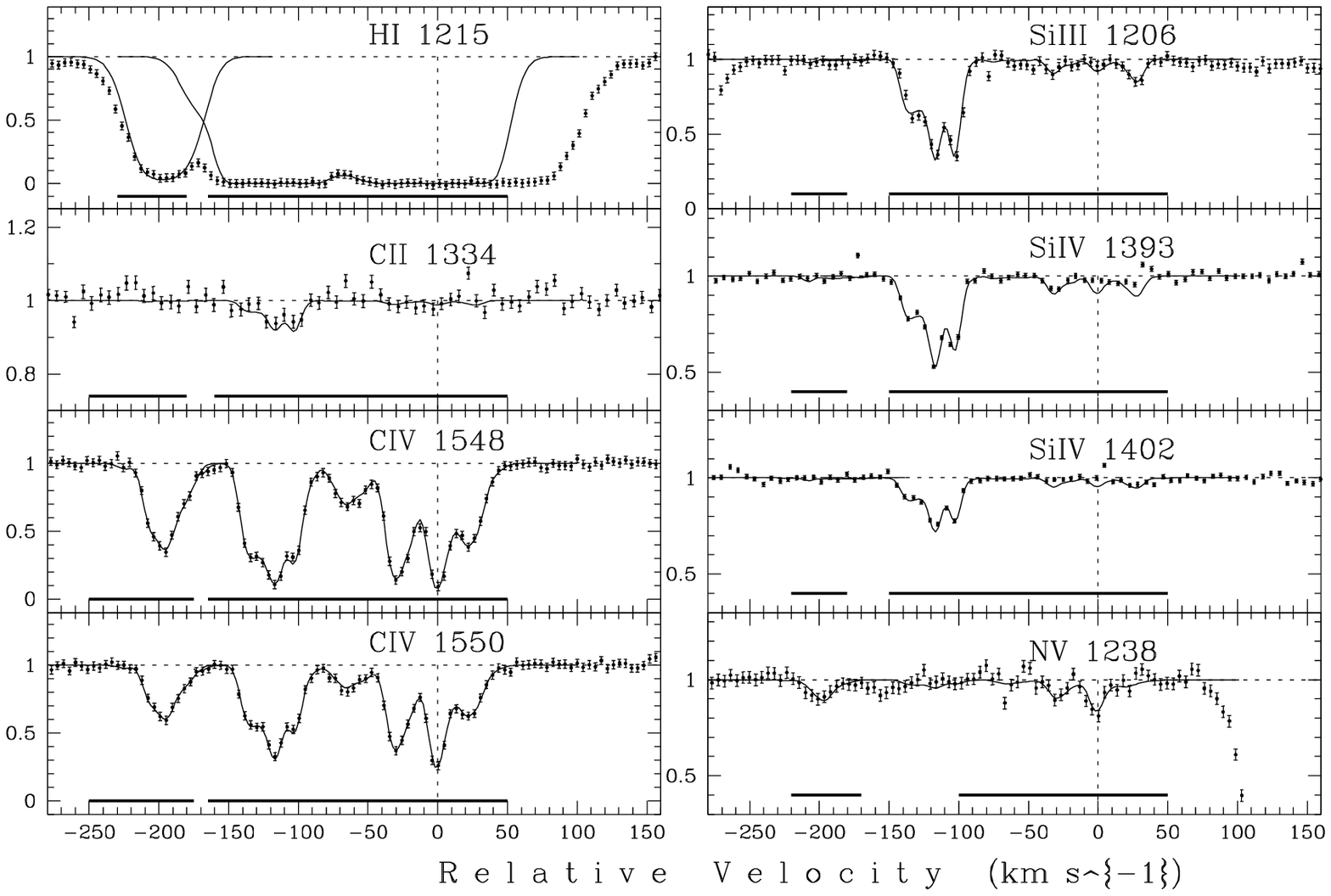,height=18.0cm,width=19.0cm}
\vspace{-7.5cm}
\caption[]{
Same as Fig.~1 but for the \zabs = 1.6838 system.
The zero radial velocity is fixed at $z = 1.68377$. 
The corresponding physical parameters are listed in Table~3, Cols.~5,6
}
\label{fig6}
\end{figure*}

In spite of all uncertainties intrinsic to this subsystem several
parameters were steadly  reproduced in all runs. These are (1)
slightly undersolar nitrogen abundance, [N/C] = $-0.1\pm0.1$, 
(2) significantly
lower ratios of Si/C and Al/C as compared to their solar values, 
[Si/C] = $-0.3\pm0.1$, [Al/C] = $-0.3\pm0.1$, and (3) extremely high
overabundance of iron [Fe/H]  $\ga 1.5$. 
The extremely high overabundance of iron as well as underabundances of
silicon and aluminium may be caused by non-equilibrium ionization.
When an absorbing gas comes to equilibrium through cooling and recombining
its ionization parameter can be overstated due to different 
cooling and recombination
times of the observed ions 
stemming from density fluctuations (so-called "hot photoionization").
A higher density gas has shorter cooling and recombination times and
hence comes faster to equilibrium. Thus, the sub-system $A$
may be described as consisting of dense 
gas clumps that are already close to equilibrium 
(seen in low ionization absorptions) 
and ambient rarefied gas still far from equilibrium
(responsible for most of \ion{C}{iv} and \ion{N}{v} absorptions).
The equilibrium ionization parameter should be lower, 
probably ranging between 0.001 and 0.003. 
With this $U_0$,  only a
mild overabundance of iron, [Fe/Si] $\sim 0.2-0.3$, 
would be enough to describe
the observed intensity of the \ion{Fe}{ii} lines.

Thus, we estimate a metallicity of $\ga 9Z_\odot$ in 
the higher ionized subsystem $B$ and
$\sim 2Z_\odot$ in the lower ionized subsystem $A$
(both metallicities are referred to silicon lines).   
This strong metallicity difference is
another argument in favor
of non-equilibrium ionization in sub-system $B$.
It probably has lower gas density  than in $A$ which leads
to longer cooling and recombination times.
We suggest that some time ago the whole absorbing complex
($A+B$) was either
exposed to a much more intense radiation
or shock-heated up to the
temperatures when collisional ionization becomes significant (Klein et al.
1994, 2003; Levshakov et al., 2004).

\subsection{Absorption system at \zabs = 1.6838}

This system separated by $\sim 12\,450$ \kms\, from the QSO shows 
a partially saturated Ly$\alpha$ and many metal lines
with complex profiles (Fig.~6). 
\ion{Si}{ii} $\lambda1260, 1193, 1190$ 
and \ion{N}{v} $\lambda1242$ are blended with
Ly$\alpha$ forest absorptions, and at the position 
of \ion{Si}{ii} $\lambda1526$ a clear continuum
window is seen.

The apparent structure of the Ly$\alpha$ profile suggests that
this system can be divided 
into two subsystems:
one at --250 \kms $< v <$ --180 \kms\, ($A$) and another 
at --180 \kms $< v < 150$ \kms\, ($B$) which
were analyzed separately. 
Calculations were carried out using both the ionizing spectrum
restored for the \zabs  = 1.7817 system and several power law spectra. 
Physical parameters
listed in Table~3, Cols.~5,6 correspond to the \zabs = 1.7817 background.
The red wing of the synthetic profile of Ly$\alpha$ in the subsystem  $B$
is calculated simultaneously with
metal lines assuming  a constant metallicity inside the absorber. 
The ionization parameter given in Table~3 for the  subsystem $A$ is determined
from the observed intensity of \ion{C}{iv} and the noise level 
at the expected position 
of the \ion{C}{ii} $\lambda1334$  line and should be considered
as a lower limit. 

The almost solar ratio of [Si/C] = $0.06\pm0.1$ and a strong
underabundance of nitrogen [N/C] $< -0.5$  obtained for
both subsystems indicate that photoionization is probably
close to equilibrium. 
However, slightly higher metallicity in the sub-system $A$ and 
the remaining flux $0.04\pm0.01$ at the 
shallow bottom of Ly$\alpha$
may indicate that hydrogen has not yet reached its equilibrium
with the ionizing background (see sub-system $B$ in the foregoing section).
Although the absolute values of the abundances are uncertain (they depend on the
assumed background and the mean ionization parameter), 
the solar to oversolar carbon content,
the ratio
[Si/C] $\simeq 0$  and a significant underabundance of nitrogen are
constantly reproduced in all trials.

\subsection{Absorption systems  at \zabs = 1.7365 and 1.4978}

Other systems with unusually strong and complex \ion{C}{iv} profiles
are identified in the spectrum of \object{HE 0141--3932}. 
Unfortunately, they cannot be analyzed by the MCI because their 
hydrogen lines are unavailable, and we describe them only qualitatively.
The system at \zabs = 1.7365 (separated by
7\,000 \kms from the QSO) shows also a clear \ion{N}{v} doublet 
(Ly$\alpha$ profile is blended).
The apparent ratio \ion{C}{iv}/\ion{N}{v}
is very much like that in the sub-system $A$
at \zabs = 1.6838 and the \zabs = 1.7365 system probably has similar 
physical parameters.
The system at \zabs = 1.4978 ($\Delta v = 30\,000$ \kms) exhibits 
a strong and complex \ion{Si}{iv} absorption
and a weak \ion{N}{v} doublet as well as \ion{C}{ii}$\lambda 1334$ 
(L$\alpha$ is out of range),
and in these features 
resembles the sub-system $B$ at \zabs = 1.6838.

\begin{table*}[t]
\centering
\caption{
Physical parameters of the \zabs = 1.7817, 1.7103 and 1.6838 metal absorbers
toward  \object{HE 0141--3932} (\zem = 1.80) derived by the MCI procedure  
(limits are given at the 1~$\sigma$ level)
}
\label{tbl-t3}
\begin{tabular}{lccccc}
\hline
\noalign{\smallskip}
 &\zabs=1.7817&\multicolumn{2}{c}{\zabs=1.7103}&
\multicolumn{2}{c}{\zabs=1.6838}\\
Parameter &  & subsystem $A$ & subsystem $B$  &subsystem  $A$  & subsystem $B$ \\
(1)$^f$ & (2) & (3)$^d$ & (4)$^d$ & (5) & (6) \\
\noalign{\smallskip}
\hline
\noalign{\smallskip}
$U_0$& 3.1E--3$^a$ & 9.0E--3 & 3.8E--2 & $\ga$2.8E--2 & 1.3E--2$^a$\\
$N_{\rm H}$, \cm& 4.2E17$^a$ & 5.7E18 & 8.6E17 & $\ga$3.2E17 & 2.5E18\\
$\sigma_{\rm v}$, \kms & 20.8$^a$ & 55.0 & 21.0 & 19.2 & 59.0$^a$\\
$\sigma_{\rm y}$& 0.63$^a$ & 0.77 & 0.62 & 0.36 & 0.52$^a$\\
$Z_{\rm C}$&3.4E--4$^a$ & 1.1E--3 & 2.1E--3 & $\la$5.2E--4 & 2.9E--4$^a$ \\
$Z_{\rm N}$&$<$1.2E--4 & 3.1E--4 & 3.2E--4 & $\la$5.0E--5 & 2.8E--5$^b$\\
$Z_{\rm Mg}$&7.0E--5$^b$ & 1.4E--4 & $\ldots$ & $\ldots$ & $\ldots$\\
$Z_{\rm Al}$&7.0E--6$^b$ & 6.2E--6 & $\ldots$ & $\ldots$ & $\ldots$ \\
$Z_{\rm Si}$&4.3E--5$^a$ & 7.4E--5 & 3.0E--4 & $\la$8.5E--5 & 4.8E--5$^a$\\
$Z_{\rm Fe}$& 6.0E--5$^b$ & $\sim$8.0E--4 & $\ldots$ & $\ldots$ & $\ldots$\\
$[Z_{\rm C}]$&$0.14\pm0.08$ &0.65 & 0.94 & $\la0.33$ & $0.08\pm0.10$\\
$[Z_{\rm N}]$&$<$0.15 & 0.56 & 0.58 &$\la-0.3$ & $-0.48\pm0.15$\\
$[Z_{\rm Mg}]$&$0.30\pm0.10$ & 0.6 & $\ldots$ &$\ldots$ & $\ldots$\\
$[Z_{\rm Al}]$&$0.30\pm0.10$ & 0.3 & $\ldots$ & $\ldots$ & $\ldots$\\
$[Z_{\rm Si}]$&$0.09\pm0.08$ & 0.33 & 0.94 & $\la$0.4 & $0.14\pm0.10$\\
$[Z_{\rm Fe}]$&$0.30\pm0.15$ & $\sim$1.5 & $\ldots$ & $\ldots$ & $\ldots$\\
$N$(H\,{\sc i}), \cm&$(2.6\pm0.5)$E15 & 1.0E16$^e$ & $(2.6\pm0.5)$E14 &
$(1.1\pm0.2)$E14 & 2.5E15$^e$\\
$N$(C\,{\sc ii}), \cm&$(1.64\pm0.08)$E13 & $(1.5\pm0.2)$E14 & $\la$1.0E12 &
$\la$1.6E11 & $(7.3\pm1.5)$E12 \\
$N$(C\,{\sc ii}$^\ast$), \cm&$<$5.0E11 & $<$4.3E12 & $\ldots$ &
$\ldots$ & $<$1.5E12 \\
$N$(Mg\,{\sc ii}), \cm&$(1.20\pm0.20)$E11 & $(5.3\pm1.5)$E12 & $\ldots$ &
$\ldots$ & $\ldots$ \\
$N$(Si\,{\sc ii}), \cm&$(2.8\pm0.3)$E12 & $(1.2\pm0.1)$E13 & $\ldots$ &
 $\ldots$ & $\ldots$\\
$N$(Si\,{\sc ii}$^\ast$), \cm&$<$1.3E11 & $<$2.0E11 & $\ldots$ & 
$\ldots$& $\ldots$ \\
$N$(Fe\,{\sc ii}), \cm&$(1.4\pm0.4)$E11 & $(5.5\pm2.0)$E11 & 
$\ldots$ &$\ldots$&$\ldots$\\
$N$(Al\,{\sc iii}), \cm &$(3.6\pm1.0)$E11 & $(1.4\pm0.3)$E12 & $\ldots$ &
$\ldots$ & $\ldots$\\
$N$(Si\,{\sc iii}), \cm &7.5E12$^c$ & $(5.5\pm0.6)$E13 & $(2.3\pm1.2)$E11 &
$\la$8.5E10 & $(7.9\pm0.8)$E12\\
$N$(C\,{\sc iv}), \cm &$(2.2\pm0.2)$E13 & $(1.7\pm0.2)$E15 & $(3.4\pm0.3)$E14 &
$(3.4\pm0.2)$E13 & $(2.3\pm0.1)$E14 \\
$N$(Si\,{\sc iv}), \cm &$(4.5\pm0.4)$E12 &$(7.7\pm0.8)$E13 & $(1.6\pm0.5)$E12 &
$<$3.0E11 & $(1.2\pm0.1)$E13 \\
$N$(N\,{\sc v}), \cm &$<$1.4E12 &$(1.8\pm0.2)$E14 & $(7.7\pm0.8)$E13 &
$(4.5\pm0.5)$E12 & $(1.3\pm0.1)$E13 \\
$\langle T \rangle$, K & 0.8E4 & 1.2E4 & 1.3E4 & 1.8E4 & 1.6E4\\
\noalign{\smallskip}
\hline
\noalign{\smallskip}
\multicolumn{6}{l}{$^{a,b}$Uncertainties: $^a$ $\sim 15-20$\%, 
and $^b$ $\sim 30$\%.}\\
\multicolumn{6}{l}{$^c$The value of $N$(\ion{Si}{iii}) is calculated 
on base of other lines.}\\
\multicolumn{6}{l}{$^d$Physical parameters ($U_0, N_{\rm H}, \sigma_{\rm v}, 
\sigma_{\rm y}$ and
metallicities) are only illustrative since estimated for the}\\ 
\multicolumn{6}{l}{\hspace*{0.2cm}photoionization equilibrium whereas 
the absorption system is in non-equilibrium.}\\
\multicolumn{6}{l}{$^e$Estimated under the assumption of constant 
metallicity inside the absorber.}\\
\multicolumn{6}{l}{$^fZ_{\rm X} = N_{\rm X}/N_{\rm H}$; 
$[Z_{\rm X}] = \log (N_{\rm X}/N_{\rm H}) -
\log (N_{\rm X}/N_{\rm H})_\odot$.}\\
\end{tabular}
\end{table*}

\subsection{The distance from the light source}

The oversolar metallicity obtained in the absorbers described above
place them in a class of associated systems.
Furthermore, their extremely large radial velocities
imply that they originate in gas ejected from
the circumnuclear region of the QSO/host galaxy.
This is also supported by the relatively strong emission flux in 
\ion{Fe}{ii} and \ion{Fe}{iii} lines and at the same time by
the \ion{Fe}{ii} absorptions in the associated systems.
Note that \ion{Fe}{ii} lines in 
absorbers with $N$(\ion{H}{i}) $\la 10^{16}$ \cm
are extremely rare; probably this is the first such detection.

The distance between the absorbing cloud and the light source
can be estimated from
the photoionization model and
the column density ratios of \ion{C}{ii}$^\ast$/\ion{C}{ii} or
\ion{Si}{ii}$^\ast$/\ion{Si}{ii}.  
For an ion in the interstellar (intergalactic) medium, the ratio
of excited to ground-state population is equal to the ratio of the
collisional excitation rate $Q_{1\rightarrow2}$ to the spontaneous
transition probability $A_{2\rightarrow1}$ (Bahcall \& Wolf 1968):
\begin{equation}
\frac{n_2}{n_1} = \frac{Q_{1\rightarrow2}}{A_{2\rightarrow1}}\: .
\label{eq:E4}
\end{equation}
The corresponding atomic data for \ion{C}{ii}$^\ast$ and
\ion{Si}{ii}$^\ast$ are the following: 
$A_{2\rightarrow1} = 2.291\times10^{-6}$ s$^{-1}$, 
the excitation rate by collisions with electrons at $T_{\rm kin} = 10^4$~K
$q^{\rm e}_{1\rightarrow2}
\simeq 1\times10^{-7}$ cm$^3$ s$^{-1}$ for \ion{C}{ii}$^\ast$, and
$A_{2\rightarrow1} = 2.17\times10^{-4}$ s$^{-1}$, 
$q^{\rm e}_{1\rightarrow2}
\simeq 2\times10^{-7}$ cm$^3$ s$^{-1}$ for \ion{Si}{ii}$^\ast$\,
(Silva \& Viegas 2002).
Since collisions with other particles have much lower excitation rates,
we put $Q_{1\rightarrow2} = q^{\rm e}_{1\rightarrow2}\,n_{\rm e}$.

We do not detect \ion{Si}{ii}$^\ast$ or 
\ion{C}{ii}$^\ast$ lines in the associated systems, therefore
clear `continuum windows' at the expected positions of 
\ion{C}{ii}$^\ast \lambda1335.7$ and
\ion{Si}{ii}$^\ast \lambda1264.7$\, were used to
set upper limits on the column densities of
\ion{C}{ii}$^\ast$ and \ion{Si}{ii}$^\ast$ (Table~3).
For the 3~$\sigma$ upper limits on  
$N$(\ion{Si}{ii}$^\ast$) and $N$(\ion{C}{ii}$^\ast$), eq.(\ref{eq:E4}) provides
$n_{\rm e} < 3$ \cmm 
in the \zabs = 1.78, and 1.71~($A$) systems, and
$n_{\rm e} < 15$ \cmm in the \zabs = 1.68~($B$) system. 
Since the degree of ionization in these systems is 
high ($n_{{\rm H}^+}/n_{\rm H} \gg 1$), the upper limits on
the total gas density $n_{\rm H}$ are the same, i.e.,
$n_{\rm H} < 3$ \cmm and $n_{\rm H} < 15$ \cmm, respectively 
(the contribution of
the ionized helium is ignored since it has a small effect). 

To estimate the distance, the QSO continuum luminosity
at the Lyman limit ${\cal L}_{\nu_c}$ must be known.
Absolute spectrophotometry is not available for \object{HE 0141--3932}. 
Therefore we estimated
the intrinsic luminosity ${\cal L}_{\nu_c}$
from ($i$) the comparison of the QSO $B$ magnitude
(assuming the QSO spectral energy distribution is a MF-type, i.e.,
$\alpha = 0.5$ in the range 912 \AA $< \lambda <$ 1600 \AA)
with the
specific flux of a star having $m_{\rm B} = 0.0$ outside the Earth's atmosphere
($4.4\times10^{-20}$ erg s$^{-1}$ cm$^{-2}$ Hz$^{-1}$) and from ($ii$)
the empirical formula given by Tytler [1987, Eq.(17)] 
which is obtained from a fit to $(f_\nu, m_v)$ data on 
over 60 QSOs with a wide range of redshifts.
The observed $B$ magnitude of 16.09 (corrected for extinction)
translates to the flux
$f_\nu$(4400~\AA) = $1.6\times10^{-26}$ erg s$^{-1}$ cm$^{-2}$ Hz$^{-1}$
which in turn leads to the apparent luminosity near the Lyman limit
${\cal L}_{\nu_c} \sim 8\times10^{31}$ erg s$^{-1}$ Hz$^{-1}$
($H_0 = 71$ km~s$^{-1}$~Mpc$^{-1}$,
$\Omega_{\rm m} = 0.3$, and $\Omega_\Lambda = 0.7$ are used to calculate
the luminosity distance of 13~Gpc).
The second method gives
${\cal L}_{\nu_c} \sim 5\times10^{31}$ erg s$^{-1}$ Hz$^{-1}$
which shows that both values are in reasonable agreement. 
In the following we use the
second one since it is based on the observational data.

Given the upper limits on $n_{\rm H}$, 
the distance from the QSO to the absorbing cloud
can be calculated from the estimated ionization parameter $U$ 
which is defined as
\begin{equation}
U = \frac{Q({\rm H}^0)}{4\pi\, r^2\,c\,n_{\rm H}} =
\frac{n_{\rm ph}}{n_{\rm H} }\; ,
\label{eq:R1}
\end{equation}
where 
\begin{equation}
Q({\rm H}^0) = \int^\infty_{\nu_c}\,\frac{{\cal L}_\nu}{h\,\nu}\,d\nu
\label{eq:R2}
\end{equation}
is the number of hydrogen ionizing photons emitted per
unit time by the central source, $c$ is the speed of light,
$\nu_c$ is the frequency of the Lyman continuum edge,
and $n_{\rm ph}$ the corresponding density of ionizing
photons.

With the estimated Lyman continuum luminosity
${\cal L}_{\nu_c} \sim 5\times10^{31}$ erg s$^{-1}$ Hz$^{-1}$,\,
one finds
$Q({\rm H}^0) \sim 7.5\times10^{57}$ photons s$^{-1}$,
assuming
${\cal L}_\nu = {\cal L}_{\nu_c}\,(\nu/\nu_c)^{-\alpha}$, and
$\alpha = 1$ in the range $\nu > \nu_c$.
A substitution of the numerical values in (\ref{eq:R1}) provides
$r_{1.68} > 100$ kpc, $r_{1.71} > 280$ kpc,
and $r_{1.78} > 450$ kpc.

\section{Discussion and conclusions}

The characteristics of both the broad emission lines and associated
absorptions of \object{HE 0141--3932} 
can be best explained by the almost pole-on view of this quasar. 
The weakness of emission lines is then due to dilution by direct
radiation from the accretion disk, whereas  
the velocity shifts of the associated systems and their distances from the
light source can be caused by the entrainment into the large-scale
outflowing jet.
However, powerful jets propagating to distances of $\sim 0.5$~Mpc
imply the existence of significant radio radiation, but
\object{HE 0141--3932} is a radio-quiet QSO.  
An upper limit to its radio flux of 1~mJy translates into
the intrinsic radio luminosity 
${\cal L}_\nu < 7\times10^{31}$ erg s$^{-1}$ Hz$^{-1}$.
This is a radio power of a typical FR~I source. Jets in the FR~I-type
structures are known to be subsonic or slightly oversonic with velocities
1\,000--10\,000 \kms, heavier than and isobaric with the external 
medium (e.g., Hughes 1991).
These characteristics are in line with the suggestion that the
observed absorptions originate in the entrained clouds.
It is also known that FR~I jets do not show an extended radio emission.
This means that radio observations of such jets, especially when they are
seen at a small angle to the line of sight,
need a high spatial resolution and high sensitivity.
Nevertheless, the calculated 
distances of several hundreds of kiloparsecs seem to be 4-5 time
overestimated since the typical jet length of the FR~I object is below
100~kpc. We cannot explain the source of this discrepancy.
It should be noted, however, 
that distances of hundreds of kiloparsecs 
for the associated systems are not exceptional 
[see, e.g., Morris et al. (1986); Tripp, Lu, \& Savage (1996); 
D'Odorico et al. (2004)].

The second fact that is hard to explain is the large velocity excess
between low- and high-ionization emission lines.
According to a model of the quasar atmosphere (e.g., Elvis 2004),
low-ionization emission
lines (H$\alpha$, \ion{O}{i}, \ion{Fe}{ii}, \ion{Mg}{ii}, \ion{C}{iii}])
come from the outer region of the accretion disk which is
well shielded from the central source radiation and optically
thick for Ly$\alpha$.
Ly$\alpha$ and high-ionization emission lines (\ion{C}{iv}, \ion{Si}{iv},
\ion{N}{v})
are formed in a cool phase of the wind arising from the inner parts of
the accretion disk.
The emission at 3400 \AA\, observed in \object{HE 0141--3932} cannot be
interpreted unambiguously, but if it is (even partially) due to
Ly$\alpha$ at $z = 1.80$, then
its redshift and strength is inconsistent with
this scenario since \ion{C}{iv} and \ion{Si}{iv} are seen at $z=1.75$.
The same situation (i.e., $z_{{\rm Ly}\alpha} > z_{\rm CIV}$)
is observed in \object{PG 1407+265}.
Apparently, a complex geometrical model of the broad
emission line region is necessary to explain the observations.

As already mentioned, the relativistic jet beamed toward us
was discovered in the bright ($B = 15.7$) and radio-quiet \object{PG 1407+265}  
(Blundell et al. 2003).
There are indications that \object{HE 0141--3932} 
may also have a similar small-scale
relativistic jet.
The Lyman limit luminosity of \object{HE 0141--3932} is  
${\cal L}_{\nu_c} = 5\times10^{31}$ erg s$^{-1}$ Hz$^{-1}$, whereas
its radio luminosity is less than
$7\times10^{31}$ erg s$^{-1}$ Hz$^{-1}$.
This implies either an unusually flat radiation continuum (typically for
QSOs is ${\cal L}_\nu \propto \nu^{-0.5}$ at $\nu < 10^{13}$~Hz and 
$\propto \nu^{-1.5}$ above), or that the apparent optical luminosity is
Doppler-boosted due to the relativistic motion of the light source.
Another argument in favor of boosting comes from the metallicity
estimations.
\object{HE 0141--3932} is a bright source. It is supposed that
the QSO luminosity is determined by the accretion rate which requires
a large amount of circumnuclear gas. This, in turn, supposes
a large mass of the contributing stellar population and, hence,
a high metal enrichment of the accreting gas. 
For reference, all QSOs with luminosities 
above $5\times10^{31}$ erg s$^{-1}$ Hz$^{-1}$
in the sample of Dietrich et al. (2003)
have metallicities $Z > 3Z_\odot$ with the mean value of 4-5$Z_\odot$.
The metallicity of 4-5$Z_\odot$ has been also measured in the associated
system of a very bright QSO \object{HE 0515--4414} (Levshakov et al. 2003b).
Gas in \object{HE 0141--3932} 
has metallicity $Z \approx 1-2Z_\odot$, which supposes
a relatively low stellar population involved in the enrichment, low
accretion rate and, hence, low luminosity.
Boosting can be caused by a relativistic jet seen at a small viewing angle.
Taking into account the measured equivalent widths of the emission lines,
we may assume that the luminosity is amplified by $\sim 10$ times,
i.e. the intrinsic luminosity of \object{HE 0141--3932} might be only
${\cal L}_{\nu_c} \sim 5\times10^{30}$ erg s$^{-1}$ Hz$^{-1}$. 

Another issue of interest is the relative abundances
obtained for the absorbing gas.
The analyzed absorption systems show high iron content, 
[Fe/C] = $0.15\pm0.1$, [Fe/Mg] = $0.0\pm0.1$ (\zabs = 1.78), 
but at the same time nitrogen
is strongly underabundant, [N/C] $\la -0.5$ (\zabs = 1.68).
Although these values were estimated in different absorption systems
they are representative for the bulk of circumnuclear gas for the
following reason.
The mass of the stellar population involved in the 
enrichment of a quasar's circumnuclear region is  
$> 10^4 M_\odot$ (Baldwin et al. 2003a) and hence large metallicity
gradients and sharp discontinuities 
due to enrichment by only a few stars
are unlikely.
In the Hamann \& Ferland (1993, 1999) models of QSO chemical evolution,
solar metallicity is reached after $\ga 0.2$ Gyr and is characterized by
a relative overabundance of nitrogen,
[N/C] $\ga 0$, and an underabundance of iron, [Fe/C] $< 0$.   
Due to the delay of 1~Gyr in Fe enrichment expected from 
the longer evolution of SNe~Ia which are the main sources of iron,
the emission line ratios \ion{Fe}{ii}/\ion{C}{iv} and 
\ion{Fe}{ii}/\ion{Mg}{ii} were proposed as a clock to constrain the QSO ages.
However, in these models, large values of [Fe/C] and [Fe/Mg]
are always associated with a considerable overabundance of nitrogen,
[N/C] $> 0.3$.  
We do not observe such 
behavior in our systems and, hence, cannot confirm this `iron clock'. This
is in line with the result of 
Matteucci \& Recchi (2001), who showed that the time scale
for enrichment by SNe Ia  is not unique but
a strong function of the adopted
stellar lifetimes, initial mass function and star formation rate and can
vary by more than an order of magnitude.

\begin{acknowledgements}
The work of S.A.L. and I.I.A. is supported by
the RFBR grant No. 03-02-17522 and by the RLSS grant 1115.2003.2.
C.F. is supported by the Verbundforschung of the BMBF/DLR grant
No. 50 OR 9911 1.
S.L. acknowledges support from the Chilean {\sl Centro de Astrof\'\i sica}
FONDAP No. 15010003, and from FONDECYT grant N$^{\rm o} 1\,030\,491$.

\end{acknowledgements}

\appendix

\section{Evaluation of the spectral shape of the local UV background}

In general, the ionizing UV background is produced by the
attenuated UV flux of QSOs and/or other sources (like young galaxies)
and by the recombination radiation from intergalactic gas clouds.
To estimate the shape of the UV continuum from the observed
profiles of ions in the metal systems we use 
the response function methodology from the theory of experimental
design (see, e.g., Box, Hunter and Hunter 1978, Chapt.~15).

At first,  the shape of the UV continuum in the range $E > 1$ Ryd 
is to be parameterized
by means of $k$ variables (factors). 
For instance, the AGN-type spectrum 
can be described by a broken power-law defined by
the following factors (see Fig.~4):
$f_1$ -- the power law index (slope) 
in the range 1 Ryd$ < E < f_2$ Ryd; $f_2$ -- the coordinate of the
first fracture;
$f_3$ the slope between $f_2$  and the point of the second fracture $f_4$;
$f_5$ the value 
of the break after $f_4$ (decimal logarithm of the intensities ratio)
with the slope $f_6$;
$f_7$ the slope in the high UV range. 
Formally we fix the starting  point of the X-ray break at $E = 100$ Ryd 
and the slope after it at  $-1.5$ 
since this spectral range affects very weakly the fractional ionizations of
ions we are interested in.
In this factor space, the Mathews-Ferland (MF) spectrum is represented by a
point with the coordinates 
$\{-0.5, 0.24, -1.0, 0.61, -2.5, -3.0, -0.7 \}$. 

At the start of the procedure we select
a basic UV spectrum (e.g. power law, HM, or MF) and
carry out the MCI calculations using the fractional
ionizations $\Upsilon_{{\rm a},i}$\footnote{If we define $n_{{\rm a},i}$
as the total ion number density of element `a' in the $i$th ionization
stage ($n_{\rm a} = \sum_i n_{{\rm a},i}$), then
$\Upsilon_{{\rm a},i} = n_{{\rm a},i}/n_{\rm a}$ is the fractional ionization
of ion $\{$a,$i\}$. For a given metallicity $Z_{\rm a}$,
$n_{{\rm a},i} = \Upsilon_{{\rm a},i} Z_{\rm a} n_{\rm H}$.}
produced by this background.

If the trial
spectrum turns out to be inconsistent with the observed line intensities, we
assign its parameters to a
`null point' in the factor space and vary
the factors around this null point according to the chosen set
of treatments\footnote{The treatment is a  particular combination
of values of the factors involved in an experiment.
The set of treatments is called the `experimental plan'.
Experimental plans for any number of factors can be found in corresponding
reference books (see, e.g., Johnson \& Leone 1977).
}. Thus, for each treatment we obtain a new trial UV spectrum.

The fitness of a particular spectrum is measured by the
value of the response function ${\cal R}$
which is defined individually for
each absorption system in such a way as
to ensure the self-consistent description
of all absorption lines detected in the system.
The best UV spectrum is that which provides
a maximum value of ${\cal R}$
within the constraints $\chi^2 \la 1$ (per degree of freedom) for each
individual line.
The information needed to construct the response function
is obtained in several test runs with the `null spectrum'.

The following example illustrates this step of the procedure.
Let us assume that an absorption system exhibits lines of
\ion{Si}, \ion{Si}{iv} and \ion{C}{ii}, \ion{C}{iv}. Trial calculations
with a basic UV continuum show that this continuum overpredicts the
intensity of \ion{Si}{ii} and underpredicts \ion{Si}{iv} whereas
the behavior of carbon lines is opposite -- underpredicted \ion{C}{ii}
and overpredicted \ion{C}{iv}. In this case it is conceivable to search
for such a background that will produce a maximum value of the product
\ion{Si}{iv}/\ion{Si}{ii} and \ion{C}{ii}/\ion{C}{iv}.
The line intensities (and column densities)
are proportional to the fractional ionizations and
we can calculate the response function simply as
\begin{equation}
{\cal R} = \frac{ \Upsilon_{{\rm Si\,}{\scriptscriptstyle\rm IV}} }
{\Upsilon_{{\rm Si\,}{\scriptscriptstyle\rm II}} }\times
\frac{ \Upsilon_{{\rm C\,}{\scriptscriptstyle\rm II}}}
{\Upsilon_{{\rm C\,}{\scriptscriptstyle\rm IV}}} \: .
\label{eq:A1}
\end{equation}
Fractional ionizations for each trial UV background are taken from CLOUDY.
Apart from the ionizing spectrum, further input parameters for CLOUDY are
the ionization parameter $U$, metallicity and element ratios
(e.g. [Si/C], [O/C]) which are estimated in test runs.

After the response function has been evaluated at each point of the
experimental plan, factor weights (influences)
$\{ \alpha_i \}^k_{i=1}$\,
can be estimated. For this purpose a standard polynomial model
is used:
\begin{equation}
{\cal R} = \sum^k_{i=1}\,\alpha_i \hat{f}_i + \beta\, ,
\label{eq:A2}
\end{equation}
where $\hat{f}_i$ is the normalized and centered value of
$i$th factor, $\hat{f}_i = (f_i - f_{0,i})/\sigma_i$,
$\sigma_i$ -- variability level of $i$th factor.
Note that this form of ${\cal R}$ defines a response surface
in $k$-dimensional factor space.
The coefficients $\alpha_i$ and $\beta$ are estimated from
$n$ ($n \geq k+1$) values of ${\cal R}$ computed at the plan
points. The input of the nonlinear term can be deduced
through the comparison of the estimate of $\hat{\beta}$
and the value of ${\cal R}_0$ at the `null point', i.e.
when $f_i = f_{0,i}$.
The statistical
significance of (${\cal R}_0 - \hat{\beta}$) points to
non-negligible factor interactions.
Then additional points (treatments) should be added to
the experimental plan to estimate the second-order terms.
The optimal spectrum is obtained by moving the factor values
in the direction normal to the response surface.
For the absorption systems described in the present paper a linear model
turned out to be sufficient.

After the first iteration of the spectral shape adjustment is completed,
we repeat the MCI calculations with the designed UV background
and, if necessary, go back to the adjustment procedure
until a satisfactory result (i.e. normalized $\chi^2 \la 1$ for
all profiles included in the analysis) is achieved. 

\end{document}